\documentclass[12pt,aps,aip,preprint]{revtex4-1}
\usepackage{graphicx}
\usepackage{amssymb}
\usepackage{subfigure}
\begin{document}

\title{Effect of Thermal Annealing on Boron Diffusion, Micro-structural, Electrical and Magnetic properties of Laser Ablated CoFeB Thin Films}

\author{G. Venkat Swamy$^1$, Himanshu Pandey$^2$, A. K. Srivastava$^1$, M.
K. Dalai$^1$, K. K. Maurya$^1$, Rashmi$^1$, and R. K. Rakshit$^{1,
}$\footnote[1]{rakshitrk@nplindia.org}}

\affiliation{ $^1$National Physical Laboratory, Council of Scientific and Industrial Research, New Delhi-110012, India.\\
$^2$Condensed Matter - Low Dimensional Systems Laboratory,
Department of Physics, Indian Institute of Technology,
Kanpur-208016, India}

\date{\today}

\begin{abstract}

We report on Boron diffusion and subsequent crystallization of
Co$_{40}$Fe$_{40}$B$_{20}$ (CoFeB) thin films on SiO$_2$/Si(001)
substrate using pulsed laser deposition. Secondary ion mass
spectroscopy reveals Boron diffusion at the interface in both
amorphous and crystalline phase of CoFeB. High-resolution
transmission electron microscopy reveals a small fraction of
nano-crystallites embedded in the amorphous matrix of CoFeB.
However, annealing at 400$^\circ$C results in crystallization of
CoFe with \textit{bcc} structure along (110) orientation.
As-deposited films are non-metallic in nature with the coercivity
(H$_c$) of 5Oe while the films annealed at 400$^\circ$C are
metallic with a H$_c$ of 135Oe.
\end{abstract}
\pacs{XX-XX, XX-XX-XX}

\maketitle

\section{Introduction}

(Co$_{x}$Fe$_{1-x}$)$_{80}$B$_{20}$ alloys have been under
extensive research focus due to high tunneling magnetoresistance
(TMR)and perpendicular magnetic anisotropy (PMA) observed in thin
films of this material combined with ultrathin MgO layer\cite{S.
Ikeda1,W. G. Wang1,W. G. Wang2,S. Ikeda2}.  A controlled
transition of CoFeB from amorphous to crystalline phase is a
necessary condition for the observation of giant TMR
effect\cite{S. Ikeda1,W. G. Wang1,W. G. Wang2}. By post deposition
thermal annealing in vacuum, TMR of the CoFeB based magnetic
tunnel junctions (MTJs) increases abruptly\cite{W. G. Wang2}.
Furthermore, recent results\cite{S. Ikeda2} have shown that
although CoFeB/MgO system is widely used for in-plane-anisotropy
MTJs, it can also meet the requirements of high thermal stability
and low induced current density magnetization switching for high
performance perpendicular MTJs. Apart from this, CoFeB has lower
coercivity (H$_c$)\cite{C. Y. You}, high spin dependent
scattering\cite{K. Nagasaka}, stronger spin tunneling
effect\cite{S. Ikeda1}, and therefore, supports a higher output
signal for particular giant magnetoresistance (GMR) and TMR ratios
in CoFeB electrode based MTJs.

From the reported results, there is a sharp contradiction in
opinions on TMR ratios due to Boron diffusion at the interface
between ferromagnetic electrode and tunnel barrier. Earlier
studies\cite{J. J. Cha,S. S. Mukherjee,J. C. Read} have shown that
improvement of TMR results from Boron diffusion at the interface
of CoFeB and tunnel barrier during annealing because it forms an
energetically favorable poly-crystalline Mg-B-O layer in case of
CoFeB/MgO interface, whereas other groups\cite{J. D. Burton,T.
Miyajima,X. Kozina,H. D. Gan} have claimed that Boron enrichment
in the barrier is detrimental to TMR because it significantly
suppresses the majority-channel conductance. Furthermore,
crystallization of amorphous CoFeB during thermal annealing at the
interface has been reported to be sensitive to its TMR effect due
to enhance coherent tunneling\cite{W. G. Wang1,S Yuasa,J.
Hayakawa}. However, crystallization of CoFeB electrode and change
in the barrier properties due to Boron diffusion at the interface
during annealing have not been fully characterized and understood
till now. Therefore, it is important to study in detail of what
happens to Boron at the interface of SiO$_2$/CoFeB as a function
of the degree of crystallization with respect to its original
amorphous phase.

To date, CoFeB thin films have been grown by using different
sputtering methods such as dc, rf and Ion beam
sputtering\cite{Shinji Yuasa,M. Raju}. Electron beam evaporation
of CoFeB thin film has also been employed successfully\cite{J. J.
Cha}. However, there are no reports on pulsed laser deposited
(PLD) thin films and junctions. The electron energy-loss
spectroscopy and x-ray photo electron spectroscopy of
CoFeB/MgO/CoFeB MTJs reveals that the existence of Boron as BO$_x$
in the barrier layer depends on the deposition method\cite{Judy J.
Cha,Abdul}. In the present study, we have explored PLD technique
as it offers a unique advantage for the growth of multi-elemental
films of desired stoichiometry of the elements with widely varying
vapour pressures. CoFeB films can be either amorphous or nano
crystalline depending on the thermal treatment, composition of the
elements and film thickness\cite{Y. M. Lee}. We carried out
Secondary ion mass spectroscopy (SIMS) depth profile measurements
to investigate the diffusion of constituent elements especially
the Boron. High-resolution transmission  electron microscopy
(HRTEM) observations reveal that CoFeB crystallized upon annealing
at 400$^\circ$C. The present work also focuses on to correlate
annealing effects on micro-structural, electrical transport and
magnetic properties of CoFeB thin films grown on amorphous layer
of SiO$_2$ coated on (001) Si substrate.

\section{Experimental}
A KrF excimer laser that produces laser pulses of width
$\approx$20 ns and wavelength of 248 nm operated at a repetition
rate of 10 Hz was used to ablate a stoichiometric target of
Co$_{40}$Fe$_{40}$B$_{20}$(at\%). The intensity of laser plume
produced an aerial energy density of 4 Jcm$^{-2}$/pulse. The PLD
deposition chamber was evacuated to the base pressure of
2x10$^{-7}$ mbar prior to deposition. The deposition pressure was
kept at 2.5x10$^{-3}$ mbar. 6N purity Ar gas was used as a buffer
gas. Under these conditions, a deposition rate of 0.4
{\AA}s$^{-1}$ was realized. The films were annealed immediately
after the deposition in the same chamber at 200, 300 and
400$^\circ$C for 1 hour under high vacuum. Our PLD grown films are
highly consistent and reproducible. Crystallographic structure of
the films was characterized using a PANalytical X'Pert PRO MRD
X-ray diffractometer with CuK$\alpha$$_1$ radiation. Elemental
analysis was carried out on a Rigaku ZSX Primus II Wavelength
Dispersive X-ray Fluorescence Spectrometer (WD-XRF). The
stoichiometry of the film as estimated by WD-XRF analysis is
within the correct stoichiometry of the target. HRTEM experiments
were carried out by employing a FEI Tecnai G2 F30 STWIN field
emission gun supported 300 kV microscope. The depth profiled data
was obtained by using TOF-SIMS 5 (ION-TOF GmbH, Germany).
Resistivity measurements were carried out in the standard four
probe geometry over a temperature range of 5-300K in a closed
cycle refrigerator. Magnetization of the samples was measured at
room temperature by using EV7 model Vibrating Sample Magnetometer
(VSM).

\section{Results and discussions}
\subsection{\textit{Structural studies}}
Grazing incident angle X-ray diffraction (GIXRD) patterns of the
as-deposited and annealed films of 40 nm thickness are shown in
Fig.1. The absence of any Bragg peak in the GIXRD scans suggests
that the as-deposited and the films annealed up to 300$^\circ$C
are amorphous in nature. But the samples annealed at 400$^\circ$C
shows a peak at 2$\theta$ = 45.07$^\circ$ which is attributed to
CoFe with bcc structure along (110) orientation. A broad diffused
peak centered at around 65.7$^\circ$ shows an evidence of the
(200) reflection. Our HRTEM results also show similar kind of
reflection at atomic scale in both as-deposited and annealed
samples. The average crystallite size of the films is determined
by the Scherrer formula is about 20 nm for annealed films at
400$^\circ$C. Strain induced in the film is calculated by
Williamson-Hall method using the formula: $\beta \cos \theta  =
\frac{{\kappa \lambda }}{\tau } + \eta \sin \theta$, where $\beta$
is the full width at half-maximum in radians, $\theta$ is the
Bragg diffraction angle of the peak, $\kappa$ is the Scherrer
constant, $\lambda$ is the wavelength of the X-rays and $\tau$ is
the crystallite size estimated from HRTEM results. The films have
tensile strain of -2.44x10$^{-4}$ generated due to vacancy-type
imperfections known as Schottky defects in the lattice created
presumably by leaving Boron into the interface.

\subsection{\textit{SIMS depth profiles}}
A uniform distribution of Co and Fe is clearly seen across the
thickness of as-deposited and annealed film (Fig.2). The peak of
Boron intensity profile (black in color) indicates strong Boron
segregation at the apex of the interface of SiO$_2$/CoFeB. It is
interesting to note that while sputtered CoFeB films suggests the
presence of Boron in amorphous phase\cite{A. A. Greer,S. V.
Karthik}, our PLD grown films in as-deposited state confirms that
Boron segregates and resides at the interface leaving CoFe in
amorphous phase. The Si signal starts rising at around 30 nm and
becomes constant at 50 nm which indicates an out-diffusion of Si
into the film structure. The Cobalt, Iron and Boron signals tends
to reduce after 40 nm which is typically the film thickness but
they are found to diffuse into the substrate as well. Si diffusion
into the film is comparatively less and it becomes constant at 44
nm in the annealed film. The diffusion of Cobalt, Iron and Boron
into the substrate is observed to be more in the annealed sample
as it is clearly visible in the depth profile. The reported
results on first principle calculations\cite{J. D. Burton} stated
that rather than inside the CoFe matrix, segregation of Boron at
the interface is energetically favorable which is consistent for
our PLD grown amorphous films.

The HRTEM micrograph of the as-deposited samples in general shows
a feature-less contrast of amorphous phase in Fig.3(a). In some
selective regions dispersed in the microstructure, little
crystalline zones have been noted [Figs.3(b) and (c)]. Further,
the microstructure observed in these regions reveals some
interesting ultra-fine details. In Fig.3(c), it is evident that
the crystalline regions are in selectively embedded state in the
amorphous matrix. A faint contrast of atomic planes in Fig.3(c)
further authenticates the significant fraction of amorphous phase
surrounding the crystalline phase, which is negligible compared to
large fraction of amorphous structure in the matrix. A region
surrounded by white dotted line displays a set of moir\'{e}
fringes evolved due to overlapping of ultra-thin tiny crystals
(Fig.3c). The careful measurement of inter planar spacings
(d$_{hkl}$) between the planes in Fig.3(c) yields (d$_{hkl}$) of
0.2 and 0.14 nm, which corresponds to hkl indices of (110) and
(200), respectively of a cubic cell (lattice parameter: a = 0.29
nm, space group: Pm\={3}m, PDF card no. 00-044-1433). A
corresponding fast fourier transform (FFT) recorded from these
nano-crystallites in an aggregate exhibits that the
nano-crystallites are in random orientation with respect to each
other and lead to Debye rings in reciprocal space. The presence of
the inter-planar spacings of 0.2 and 0.14 nm corresponding to the
planes (110) and (200) respectively are marked on FFT pattern
(inset in Fig.3c). A relatively more intense plane with higher
fraction compared to (200) planes was the existence of (110) set
of atomic planes in the microstructure with the inter-planar
spacings of about 0.2 nm in Figs.3(c) and (d). Lattice scale
images of (110) planes in the amorphous matrix reveals a better
grey level contrast (Fig.3d) compared to that of (200) planes
(Fig.3c).

The overall microstructure of annealed film at 400$^\circ$C as
shown in Figs.4(a) to (d) has a distinct grey level contrast
consists of several nano-crystallites of the size even up to 15 to
20 nm. These nano-crystallites marked with a set of white arrows
[Figs. 4(a) and (b)] are separated from each other with apparent
interface of the boundary length between 5 to 10 nm. It is evident
that the film microstructure is aligned preferably along the well
oriented atomic planes of (110), a cubic crystal within individual
nano-crystallites. The dominant atomic planes in the matrix are
observed as (110) although the formation of (200) planes exists
[Figs.4(c) and (d)]. A corresponding FFT [inset in Fig.4(d)]
further elucidates a set of atomic planes in reciprocal space with
inter-planar spacings of 0.29 and 0.14 nm.

\subsection{\textit{Electrical transport studies}}
Electrical transport properties of CoFeB films in their
as-deposited state and annealed at 400$^\circ$C are shown in
Fig.5. Resistivity data has been normalized to its value at 273K.
The room temperature resistivity of as-deposited films are about
$\approx $70$ \mu\Omega$cm and for samples annealed at
400$^\circ$C is observed to be $\approx $8$ \mu\Omega$cm
respectively. The abrupt increase in resistivity values for
amorphous films and observed non-metallic behavior could be due to
intragrain tunneling and the SiO$_2$ inclusions into the film was
reported previously in granular CoFeB-SiO$_2$ amorphous thin film
system\cite{P. Johnsson}. Our HRTEM results indicate the presence
of nano crystalline regions in the amorphous matrix of CoFeB,
these crystalline regions are well separated and leading to the
absence of metallic channels support the observed transport
behavior. The resistivity is found to reduce gradually on
decreasing temperature from 300K followed by a clear increase in
its value below 22K for the film annealed at 400$^\circ$C. A fit
to the part of our experimental data below the resistivity minimum
for annealed sample using the following empirical relation
${\rho(T)} ={\beta_0} + {\beta}lnT, 5K \leq T \leq 22K$ is shown
as a solid line in the figure. Clear agreement with experimental
data suggests finite magnetic contribution as has been observed in
several amorphous ferromagnetic alloys at low temperature\cite{S.
N. Kaul}. The detailed analysis of our transport measurements is
under progress.

Magnetization [M(T)] measurement performed on as-deposited and
annealed CoFeB films at different temperatures are shown in Fig.6.
The ferromagnetic hysteresis loops after substrate correction for
in-plane and out-of-plane orientations confirmed that films
exhibit in-plane easy axis of magnetization. In the annealed film,
the easy axis loop is squared shape, while the hard axis of the
film displays a much slanted curve indicating the presence of
magnetic anisotropy\cite{J. McCord}. The as-deposited samples and
samples annealed up to 300$^\circ$C shows H$_c$ of $\sim$10Oe. The
films annealed at 400$^\circ$C are found to show increase in H$_c$
from 5Oe to 135Oe. A sharp increase in H$_c$ at annealing
temperature of 400$^\circ$C is a clear evidence of crystallization
of amorphous CoFeB films.

\section{Conclusions}
In summary, we have grown thin film of CoFeB using pulsed laser
ablation of an alloy target. HRTEM together with resistivity
measurements reveal that annealing at 400$^\circ$C increases the
metallicity of the film by forming nano-crystallites. The sudden
increase in H$_c$ from 5Oe to 135Oe after annealing at
400$^\circ$C also confirms the crystallization of CoFeB. Magnetic
hysteresis loops of in-plane and out-of-plane measurements reveal
that our films have in-plane easy axis of magnetization. We
believe Boron segregation at the interface of SiO$_2$/CoFeB,
observed in our SIMS depth profiles, have important repercussions
on the TMR values.

\section{Acknowledgments}
 This research has been supported by Council of Scientific
and Industrial Research (CSIR), Government of India. The authors
are grateful to Prof. R. C. Budhani, Director NPL for his valuable
suggestions. Venkat and Himanshu acknowledge CSIR for financial
support.

\clearpage
\begin{figure}
\begin{center}
%\vskip -1cm
%\abovecaptionskip -10cm
\includegraphics [width=15cm]{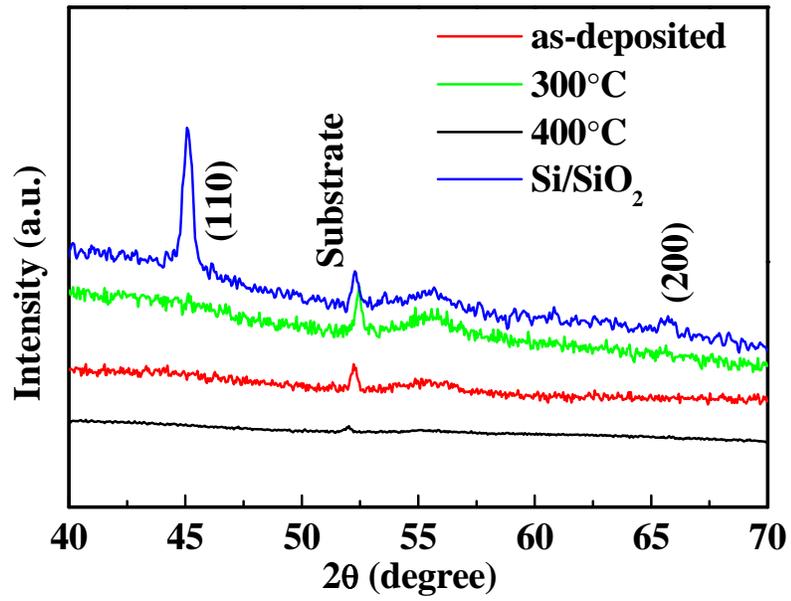}%
\end{center}
%\vskip -1.3cm
\caption{\label{Fig1} GIXRD measurements of bare substrate
(Si/SiO$_2$) and Co$_{40}$Fe$_{40}$B$_{20}$ thin film of thickness
40 nm in as-deposited and annealed states.}
\end{figure}

\clearpage
\begin{figure}
\begin{center}
%\vskip -0.8cm
%\abovecaptionskip -10cm
\includegraphics [width=15cm]{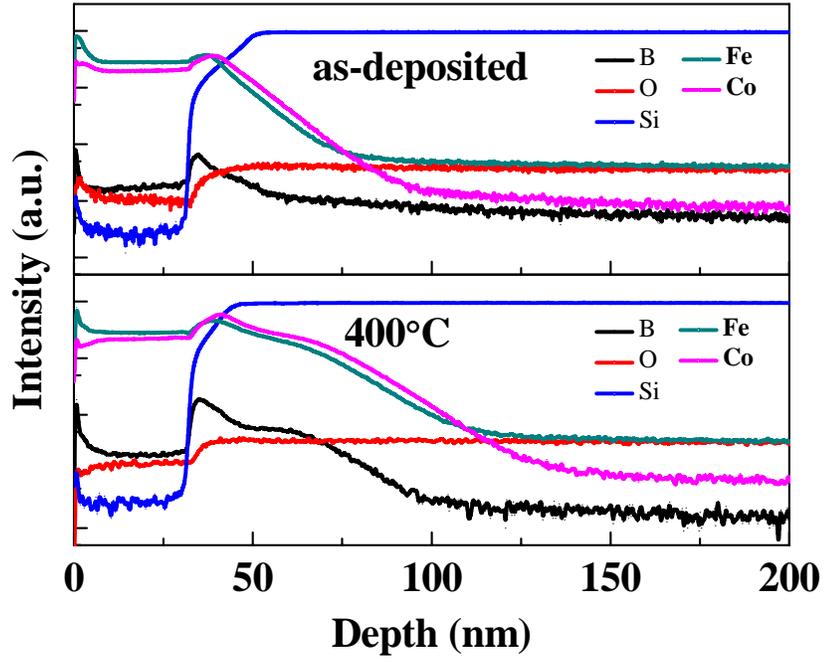}%
\end{center}
%\vskip -1.3cm
\caption{\label{Fig2} Secondary Ion Mass Spectroscopy (SIMS) depth
profiles of as-deposited and annealed CoFeB thin films of 40 nm
thick at 400$^\circ$C showing strong Boron (B) diffusion at the
interface.}
\end{figure}

\clearpage
\begin{figure}
\begin{center}
%\vskip -1.2cm
%\abovecaptionskip -10cm
$\includegraphics [width=8cm]{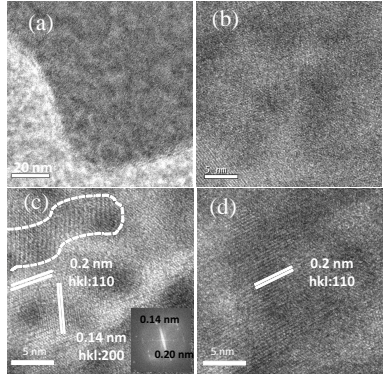}$ %
\end{center}
%\vskip -0.5cm
\caption{\label{Fig3} HRTEM micrographs of CoFeB as-deposited film
showing (a) amorphous phase, (b - d) a small fraction of
nano-crystallites distributing in the dominated amorphous matrix.
Inset in (c) shows a corresponding FFT in reciprocal space.}
\end{figure}

\clearpage
\begin{figure}
\begin{center}
%\vskip -1.2cm
%\abovecaptionskip -10cm
\includegraphics [width=8cm]{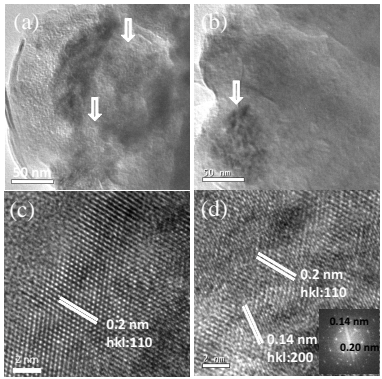} %
\end{center}
%\vskip -0.5cm
\caption{\label{Fig4} HRTEM micrographs of CoFeB film annealed at
400$^\circ$C showing (a - b) crystalline microstructure, (c - d)
atomic planes of crystalline phase constituting the entire matrix
of the film. Inset (d) shows a corresponding FFT in reciprocal
space.}
\end{figure}

\clearpage
\begin{figure}
\begin{center}
%\vskip -1cm
%\abovecaptionskip -10cm
\includegraphics [width=15cm]{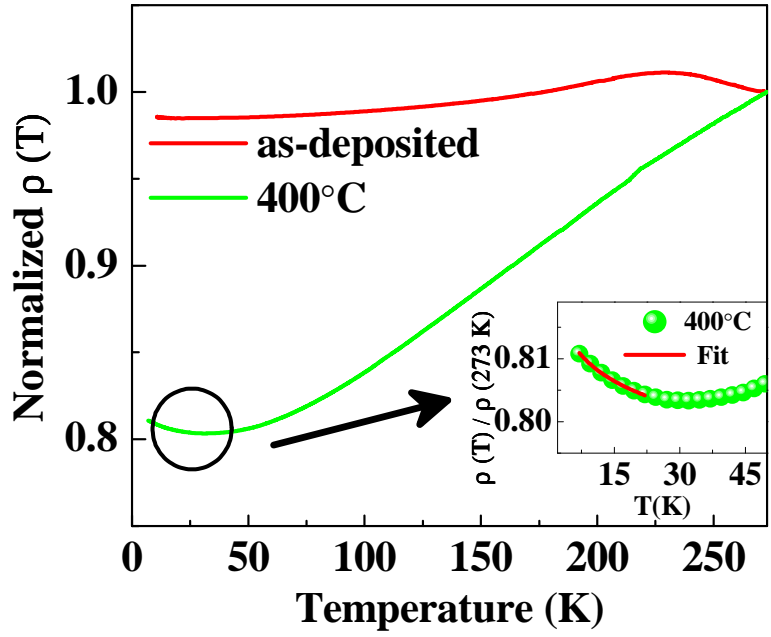}%
\end{center}
%\vskip -0.8cm
\caption{\label{Fig (5)} Normalized resistivity graph of
Co$_{40}$Fe$_{40}$B$_{20}$ as-deposited and annealed films. Inset
shows zoomed image of the electrical resistivity of 400$^\circ$C
annealed film. The solid line in the inset represents fit to the
equation in the temperature range of 7-22K.}
\end{figure}

\clearpage
\begin{figure}
\begin{center}
%\vskip -1cm
%\abovecaptionskip -10cm
\includegraphics [width=15cm]{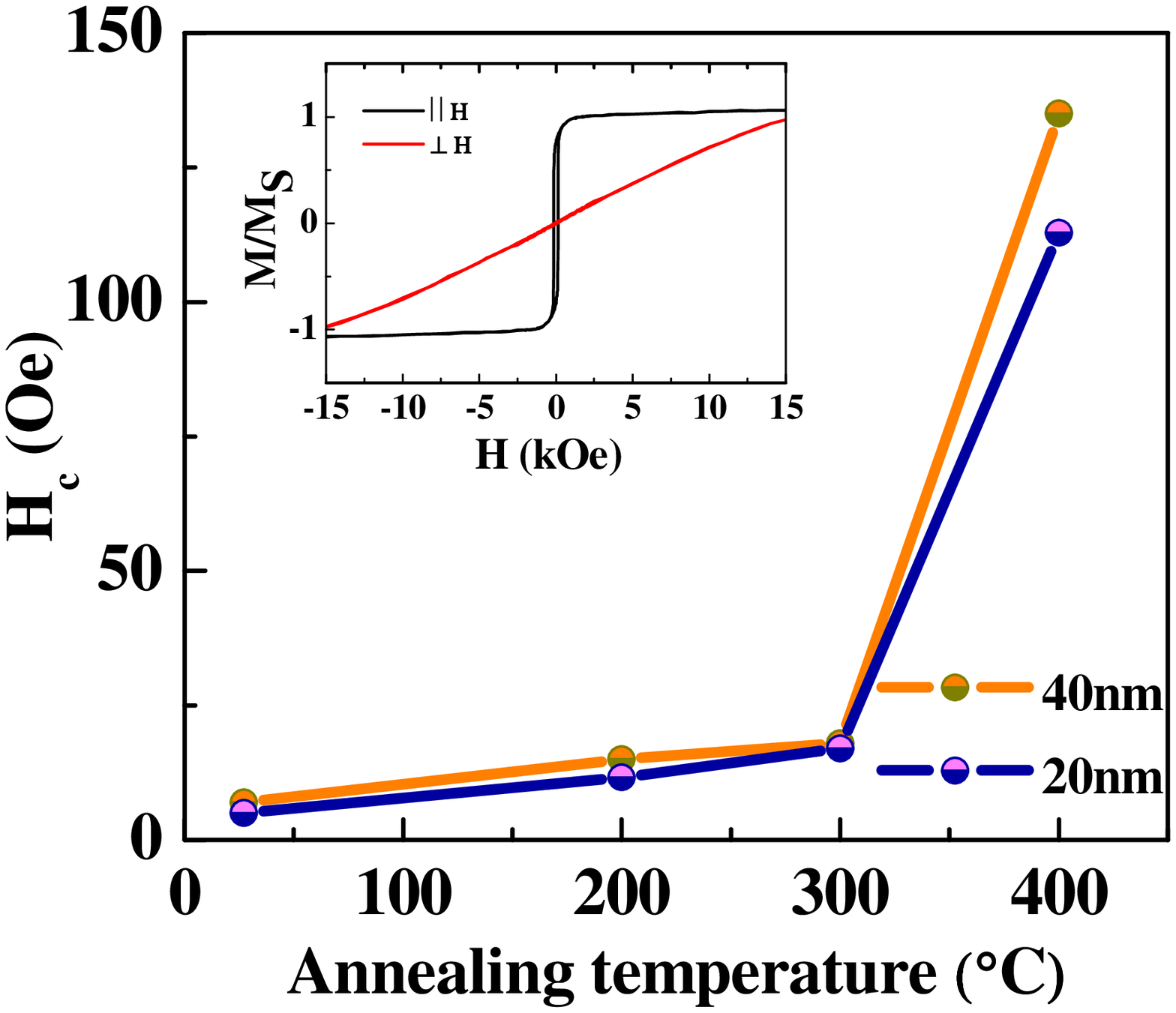}
\end{center}
%\vskip -0.6cm
\caption{\label{Fig6} H$_c$ plots as a function of annealing
temperature for CoFeB of thickness 40 and 20 nm. Inset shows the
room temperature magnetic hysteresis loops for 400$^\circ$C
annealed film measured along parallel and perpendicular field
direction with respect to film plane.}
\end{figure}

\clearpage
\section{List of Figures}

FIG.1. GIXRD measurements of bare substrate (Si/SiO$_2$) and
Co$_{40}$Fe$_{40}$B$_{20}$ thin film of thickness 40
nm in as-deposited and annealed states.\\

FIG.2. Secondary Ion Mass Spectroscopy (SIMS) depth profiles of
as-deposited and annealed CoFeB thin films of 40 nm thick at
400$^\circ$C showing strong Boron (B) diffusion at the interface.\\

FIG.3. HRTEM micrographs of CoFeB as-deposited film showing (a)
amorphous phase, (b - d) a small fraction of nano-crystallites
distributing in the dominated amorphous matrix. Inset in (c) shows
a corresponding FFT in reciprocal space.\\

FIG.4. HRTEM micrographs of CoFeB film annealed at 400$^\circ$C
showing (a - b) crystalline microstructure, (c - d) atomic planes
of crystalline phase constituting the entire matrix of the film.
Inset (d) shows a corresponding FFT in reciprocal space.\\

FIG.5. Normalized resistivity graph of Co$_{40}$Fe$_{40}$B$_{20}$
as-deposited and annealed films. Inset shows zoomed image of the
electrical resistivity of 400$^\circ$C annealed film. The solid
line in the inset represents fit to the equation in the
temperature range of 7-22K.\\

FIG.6. H$_c$ plots as a function of annealing temperature for
CoFeB of thickness 40 and 20 nm. Inset shows the room temperature
magnetic hysteresis loops for 400$^\circ$C annealed film measured
along parallel and perpendicular field direction with respect to film plane.\\

\end{document}